\documentclass[optics,article,submit,pdftex,moreauthors]{Definitions/mdpi}

\firstpage{1} 
\pubvolume{1}
\issuenum{1}
\articlenumber{0}
\pubyear{2025}
\copyrightyear{2025}
\datereceived{ } 
\daterevised{ } 
\dateaccepted{ } 
\datepublished{ } 
\hreflink{https://doi.org/} 

\Title{Infrared Imaging of Photochromic Contrast in Thiazolothiazole-Embedded Polymer Films}

\TitleCitation{Title}

\Author{Nuren Z. Shuchi $^{1,\ast}$\orcidA{}, Tyler J. Adams $^{2}$\orcidB{}, Naz F. Tumpa$^{2}$\orcidC{}, Dustin Louisos$^{1}$\orcidF, Glenn D. Boreman$^{1}$\orcidG, Michael G. Walter$^{2}$\orcidD{}, and Tino Hofmann$^{3}$\orcidE{}}

\AuthorNames{Firstname Lastname, Firstname Lastname and Firstname Lastname}

\address{%
$^{1}$ \quad Department of Physics and Optical Science, University of North Carolina at Charlotte, \newline 9201 University City Blvd, Charlotte, NC, 28223 \\
$^{2}$ \quad Department of Chemistry, University of North Carolina at Charlotte, 9201 University City Blvd, \newline Charlotte, NC, 28223 \\
$^{3}$ \quad Department of Physics, New Jersey Institute of Technology, University Heights, Newark, NJ, 07102}

\corres{Correspondence: nshuchi@charlotte.edu}

\abstract{The increasing demand for optical technologies with dynamic spectral control has driven interest in chromogenic materials, particularly for applications in tunable infrared metasurfaces. Phase-change materials such as vanadium dioxide and germanium–antimony–tellurium, for instance, have been widely used in the infrared regime. However, their reliance on thermal and electrical tuning introduces challenges such as high power consumption, limited emissivity tuning, and slow modulation speeds. Photochromic materials may offer an alternative approach to dynamic infrared metasurfaces, potentially overcoming these limitations through rapid, light-induced changes in optical properties. This manuscript explores the potential of thiazolothiazole-embedded polymers, known for their reversible photochromic transitions and strong infrared absorption changes, for tunable infrared metasurfaces. The material exhibits low absorption and a strong photochromic contrast in the spectral range from 1500 cm$^{-1}$ to 1700 cm$^{-1}$, making it suitable for dynamic infrared light control. This manuscript reports on  infrared imaging experiments demonstrating photochromic contrast in thiazolothiazole-embedded polymer and thereby provides compelling evidence for their potential applications for dynamic infrared metasurfaces.}


\begin{document}


\section{Introduction}

The growing demand for advanced optical technologies capable of dynamic manipulation of spectral properties through external stimuli has spurred significant interest in chromogenic materials with tunable optical properties. These chromogenic materials can reversibly alter their optical properties in response to external stimuli such as light, heat, electric or magnetic fields, mechanical stress, and chemical environments. \citep{granqvist1990chromogenic,Woodward2017, Sayresmith2019photostable}. Applications for these chromogenic materials include high-density optical recording systems, tunable metasurfaces, tinted lenses, smart windows, memory devices, actuators, tunable filters, and imaging \citep{Chen2005organic,Che2020tunable,Tinted_eyewear,Cho_kim_MemoryDevice,Actuator,Tunable_filter,Chen2018biochromic}. 

Among these applications, tunable optical metasurfaces are of particular interest. Metamaterials composed of engineered subwavelength structures interact strongly with light, allowing precise control over its amplitude, phase, and polarization \citep{pendry2006}. This enables design and fabrication of compact optical devices, addressing the limitations of traditional bulky components. \citep{Ding2022,huang2017reconfigurable}. Examples of metasurface applications include beam steering, planar lenses, holography, as well as narrowband and broadband filters \citep{Ding2022,Zang2018polarization,Wang2016plasmonic,Tasolamprou2022metasurface_absorber}.

In the infrared regime, phase-change materials like vanadium dioxide and germanium–antimony–tellurium have been utilized where tunability is achieved through thermally- and electrically-induced metal-insulator phase transitions that result in significant changes in their optical properties \citep{Zhang2022Vanadium,Tang2023GST}. However, these methods face several practical challenges, including high power consumption due to the need for continuous external power, a limited emissivity tuning range, and slow modulation speed \citep{xiong2024electrically}. In comparison with tuning using thermal and electrical stimuli, optical modulation of metasurface components presents a direct and often rapid modulation approach \citep{Kang2019}.

Chromogenic materials that exhibit reversible changes in their optical properties in response to optical stimuli are called photochromic materials \citep{crano1999organic,konaka2003syntheses}.
These materials have been demonstrated to play a significant role in facilitating the development of tunable infrared metasurfaces by leveraging their light-induced changes in optical properties \citep{Bang2018}. The development of photochromic materials that exhibit strong and reversible changes in their optical properties in the infrared region could offer an alternative approach to achieving tunable infrared metasurfaces, potentially with advantages in terms of cost, fabrication, or power consumption.

Viologens represent an important class of photochromic materials \cite{guo2024_Viologen}. Their properties can be enhanced by incorporating a thiazolo[5,4-d]thiazole (TTz) fused, conjugated bridge, an approach that has gained growing interest due to its strong fluorescence, solution-processability, and reversible photochromic transitions. Notably, dipyridinium TTz viologens exhibit high-contrast, rapid, and reversible photochromic changes when integrated into a polymer matrix. Upon exposure to radiation with energy exceeding 2.8~eV, they undergo a color transition from light yellow (TTz$^{2+}$) to purple (TTz$^{\cdot+}$) and then to blue (TTz$^{0}$) due to two distinct photoinduced single-electron reductions. The reverse transition occurs through the reaction of the TTz$^{0}$ state with molecular oxygen \citep{Tadams2024_TTz_2024}.

The optical properties of photochromic TTz-embedded in a polymer matrix have been reported in the visible, near-infrared and infrared spectral ranges \citep{Tadams2024_TTz_2024,shuchi2025infrared} . TTz-embedded polymer exhibits strong changes in the absorption bands in the infrared spectral range from 500~cm$^{-1}$ to 1800~cm$^{-1}$ due to the photochromic transition. In the narrow spectral range from 1500 cm$^{-1}$ to 1700 cm$^{-1}$, TTz-embedded polymer shows low absorption and a strong photochromic contrast between its states before and after irradiation. These characteristics, specifically the photochromic contrast coupled with low absorption in this mid-infrared window, maybe utilized for the development of dynamic metasurfaces. In this manuscript, the potential of TTz-embedded polymers as a promising material for dynamic infrared metasurfaces is investigated. Infrared imaging experiments demonstrating transmission contrast offer strong initial validation of their effectiveness in the infrared spectral range.
\section{Experiment}

Dipyridinium TTz was synthesized by refluxing dithiooxamide and 4-pyridine\-carboxal\-de\-hyde in dimethylformamide at 153$^{\circ}$C for 8 hours. The resulting dipyridyl TTz was then treated with 3-bromopropyl trimethylammonium bromide in dimethylformamide to alkylate the pyridine rings, enhancing its water solubility. Further details on the synthesis of dipyridinium TTz can be found in our previous publications \cite{adams2021, Tadams2024_TTz_2024}. To fabricate the TTz-embedded polymer hydrogel samples, 3.4 w\% dipyridinium TTz was dissolved in a polyvinyl alcohol solution, followed by the addition of borax, as described in previous studies \cite{adams2021}. The viscous polymer $/$ TTz mixture material was blade-coated and dried in ambient air for 24 h to obtain a nominally 100~\textmu m thick polymer film sample.

\begin{figure}[H]
\centering
\includegraphics[width=.9\textwidth]{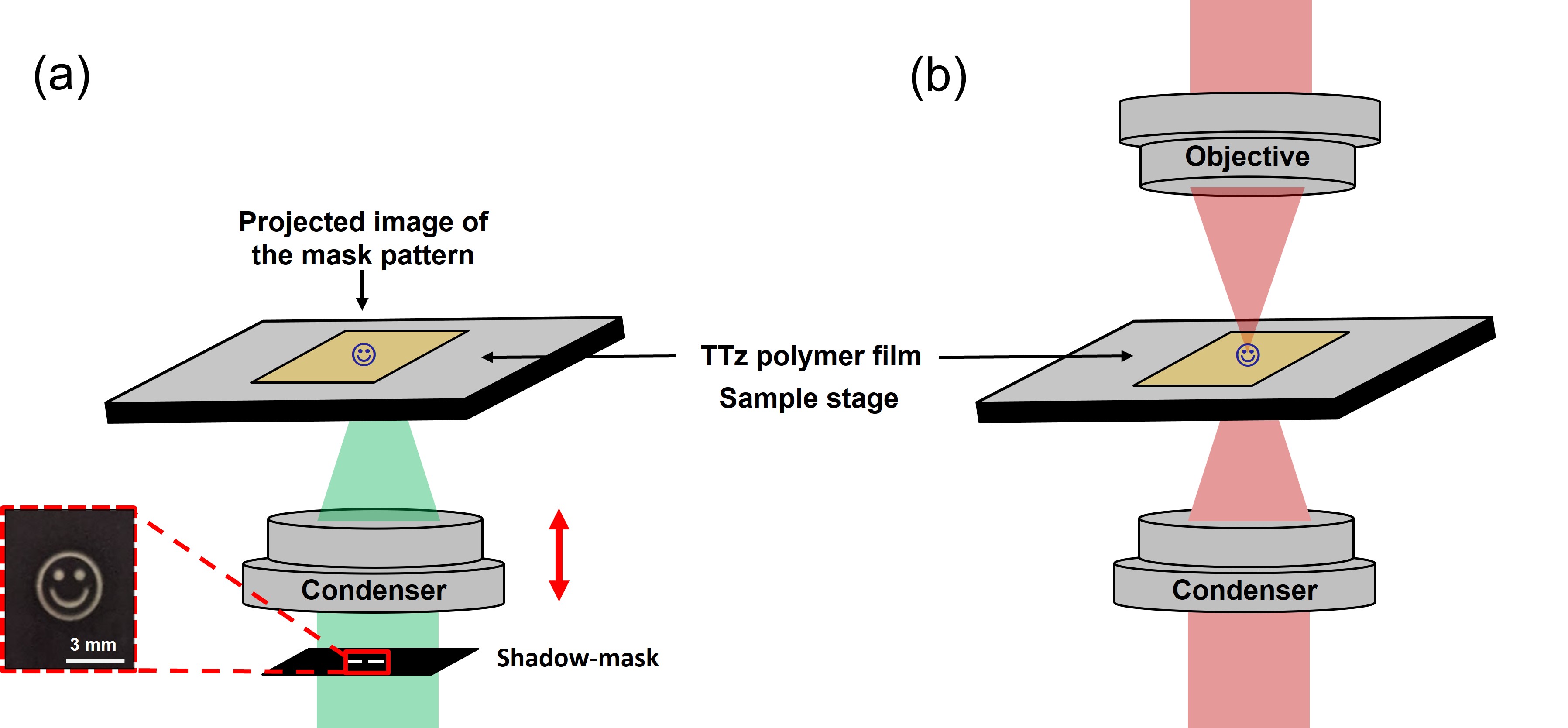}
\caption{Arrangement for exposure (a) and imaging (b). For the exposure, the broadband white-light source of the microscope is used. A shadow mask is positioned in the collimated beam path before the condenser lens, and the condenser is adjusted to project an image of the shadow mask onto the TTz-embedded polymer sample. 
After 10 sec exposure, the white light source is turned off and a KRS-5 optical filter with a cut-on wavelength of 700~nm is introduced in the source side beam path before the sample. With the optical filter in place, unintended photochromic changes induced by the broadband visible light source during visible image acquisition are prevented.}
\label{Recording and imaging}
\end{figure}

The TTz-embedded polymer sample was investigated using a Hyperion 3000 microscope (Bruker) in combination with a Vertex 70 Fourier-transform infrared spectrometer (Bruker). The Bruker Hyperion 3000 microscope utilizes a confocal beam path and allows data acquisition in both transmission and reflection modes. The microscope is equipped with a broadband LED in the visible spectral range, while a silicon carbide globar serves as the light source in the infrared spectral range. The instrument features two infrared detectors: a single element mercury cadmium telluride (MCT) detector and a 64 $\times$ 64  (4,096 pixels) MCT focal plane array (FPA). The MCT FPA allows the acquisition of hyperspectral infrared images in the spectral range from 900~cm$^{-1}$ to 4000~cm$^{-1}$. A shadow mask was employed to facilitate spatially controlled exposure of the TTz-embedded polymer films using the microscope's broadband white-light source. The patterned mask depicted in the inset of Figure~\ref{Recording and imaging}(a) was fabricated using transparency sheets printed with the desired pattern. The pattern was composed of transparent and opaque regions to allow white-light exposure of the TTz-embedded polymer sample in desired areas. 

Figure~\ref{Recording and imaging} illustrates the optical arrangement for visible light exposure and imaging in transmission mode using the microscope. As depicted in Figure~\ref{Recording and imaging}(a), the shadow mask was placed in the collimated illumination path of the microscope before the condenser lens. The vertical position of the condenser was adjusted to ensure that the shadow mask plane and the sample plane were optically conjugate. 

This configuration enabled the condenser lens to project a 6x reduced image of the mask pattern on the TTz-embedded polymer sample with high spatial fidelity. After photo-induced patterning of the TTz-embedded polymer film, the white light source was turned off and the condenser lens was re-adjusted to restore the microscope to its confocal imaging configuration. In this arrangement, the MCT FPA was employed to acquire infrared transmission images, while a commercial CCD camera array was used to record the photochromic contrast in the visible as shown in Figure~\ref{Recording and imaging}(b). The images were obtained at room temperature in ambient oxygen at four different x-y positions to capture the entire photochromic pattern using a high-accuracy, motorized x-y sample translation stage.

\section{Results and Discussion}

\begin{figure}[H]
\centering
\includegraphics[width=.85\textwidth]{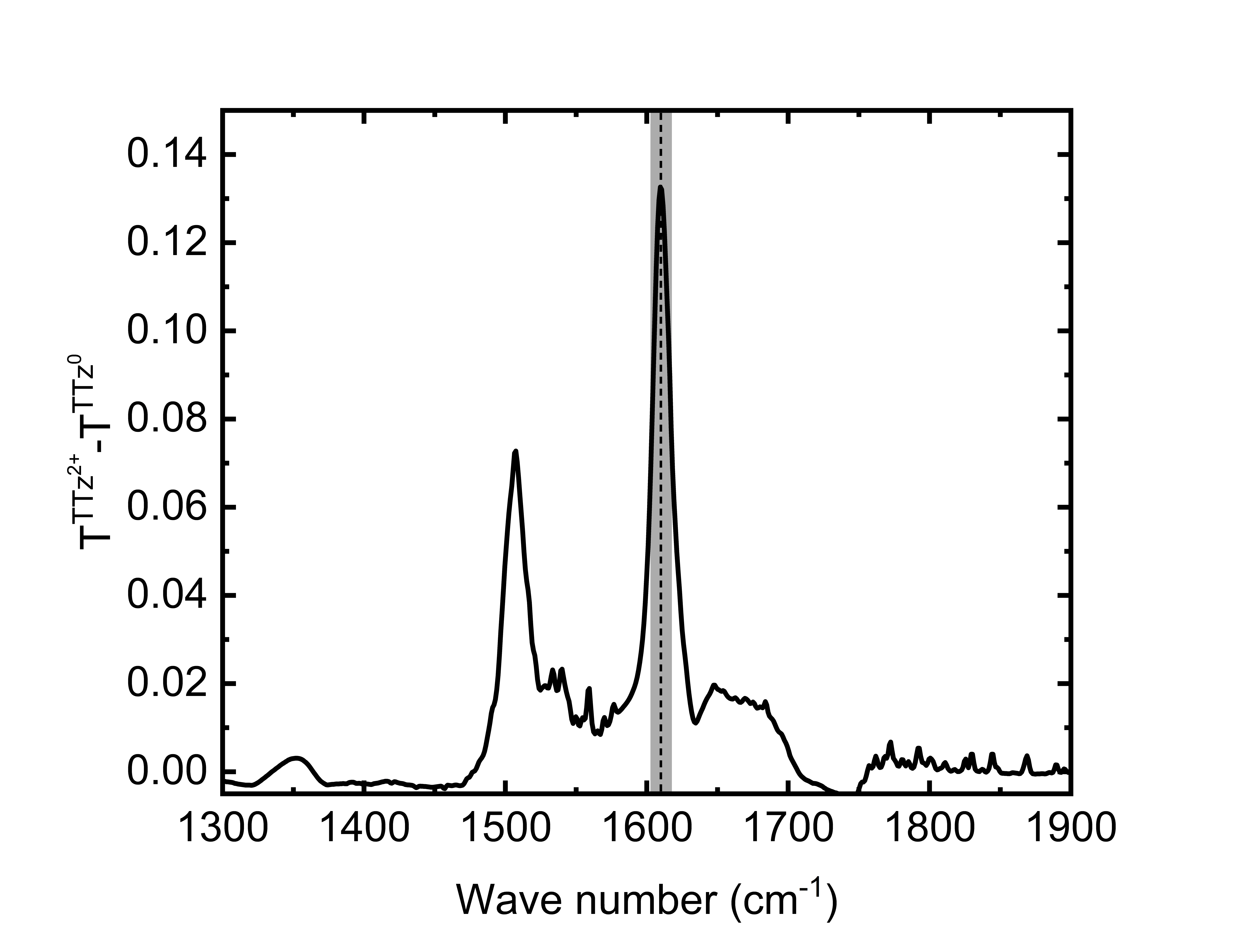}
\caption{Unpolarized transmission difference of the TTz-embedded polymer sample before (T$^{\text{TTz}^{2+}}$) and after (T$^{\text{TTz}^{0}}$) the photochromic transition in the spectral range from 1300~cm$^{-1}$- 1900~cm$^{-1}$. The maximum transmission difference, indicating the largest change in absorption due to the photochromic transition, is observed at 1610~cm$^{-1}$ indicated by the dashed vertical line. For the transmission images shown in Fig.~\ref{Vis and IR image.jpg} the intensity is integrated for the full width half maximum of the peak located at 1610~cm$^{-1}$ as indicated by the shaded area.}
\label{Transmission Spectrum}
\end{figure}

In order to ensure the acquisition of infrared images with observable photochromically-induced changes, the identification of spectral bands that exhibit adequate dielectric contrast between the TTz$^{2+}$- and TTz$^{0}$-states of the TTz-embedded polymer is required. For imaging in transmission configuration, it is further necessary to identify spectral regions that are sufficiently transparent. Spectroscopic ellipsometry measurements in the infrared spectral range revealed that in the spectral range from 1500 cm$^{-1}$ to 1700 cm$^{-1}$ the TTz-embedded polymer demonstrates low absorption and a notable dielectric contrast between its states before and after irradiation with a 405~nm diode laser source (see Fig.~3 (b) in Ref.~\cite{shuchi2025infrared}). In order to further narrow the band that could provide sufficient contrast for transmission infrared imaging, unpolarized transmission measurements were performed on the TTz-embedded polymer before ($T^{\text{TTz}^{2+}}$) and after ($T^{\text{TTz}^{0}}$) the photochromic transition in the spectral range from 1300~cm$^{-1}$- 1900~cm$^{-1}$. The transmission difference spectrum ($T^{\text{TTz}^{2+}} - T^{\text{TTz}^{0}}$) of these unpolarized transmission measurements is shown in Figure~\ref{Transmission Spectrum}.

\begin{figure}[H]
\centering
\includegraphics[width=.85\textwidth]{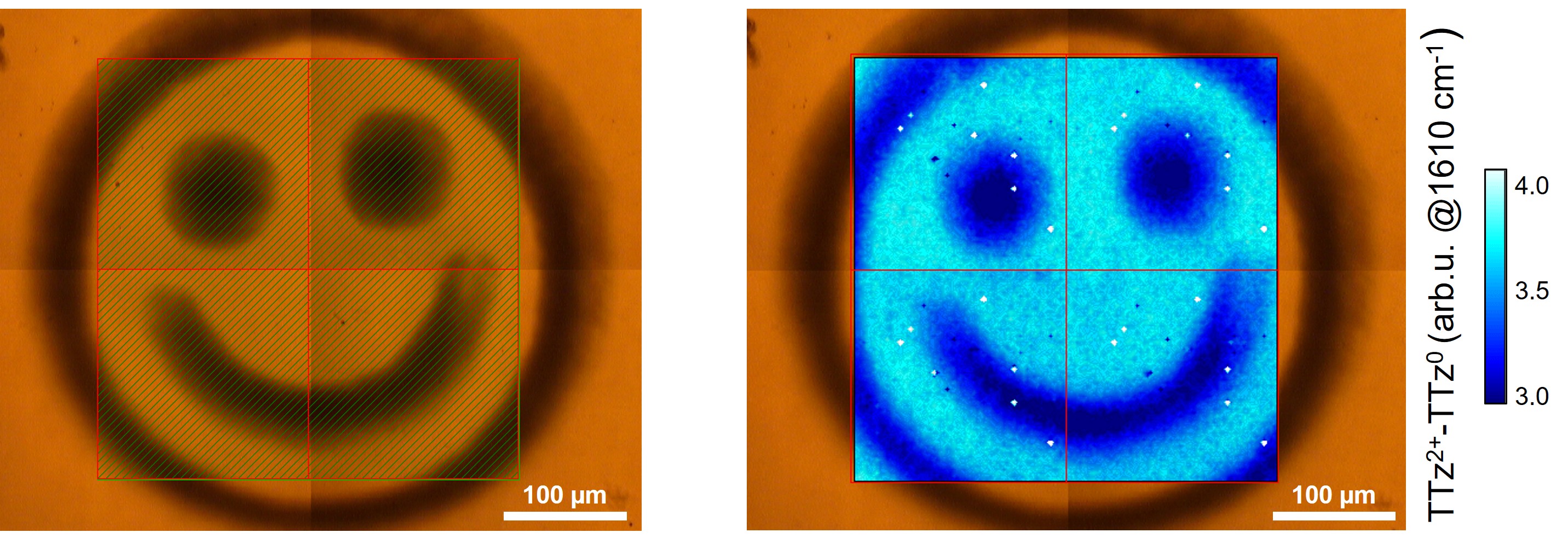}
\caption{Optical microscope images taken in transmission mode of the TTz-embedded polymer sample with the recorded pattern. The left panel depicts the transmission image in the visible spectral range,  where the background corresponds to the TTz$^{2+}$ state, while while the recorded pattern appears in the TTz$^{0}$ state. The right panel shows a false-color representation of the narrow-band infrared transmission contrast centered at 1610~cm$^{-1}$, obtained by integrating the hyperspectral imaging data of the TTz-embedded polymer over the full width at half maximum of the peak located at 1610~cm$^{-1}$. This specific band centered around 1610~cm$^{-1}$ was  selected based on the unpolarized transmission data to maximize the transmission contrast associated with the photochromic transition between the TTz$^{2+}$ and TTz$^{0}$ states. A scale bar of 100 \textmu m is included in both images for reference.}
\label{Vis and IR image.jpg}
\end{figure}

The unpolarized transmission-difference spectrum in the spectral range from 1300~cm$^{-1}$ to 1900~cm$^{-1}$ is dominated by two strong and narrow absorption lines centered around 1500~cm$^{-1}$ and 1610~cm$^{-1}$. For the demonstration of the infrared transmission imaging of photochromically-induced changes in TTz-embedded polymer, the strongest absorption line, located at 1610~cm$^{-1}$, was selected. For the false-color infrared transmission images shown in  Fig.~\ref{Vis and IR image.jpg} the area of the absorption line located at 1610~cm$^{-1}$ is integrated for the full width half maximum of the peak, i.e., from 1600~cm$^{-1}$ to 1618~cm$^{-1}$ as indicated by the shaded area in Fig.~\ref{Transmission Spectrum}.

Figure~\ref{Vis and IR image.jpg} presents imaging results for both visible and infrared spectral ranges, captured with a commercial CCD camera array and a 64 $\times$ 64 pixel MCT FPA, respectively. The left panel shows a transmission image of the TTz-embedded polymer in the visible spectral range where the background in TTz$^{2+}$ state contrasts with the recorded pattern in TTz$^{0}$ state. A KRS-5 optical filter with a cut-on wavelength of 700~nm is used during the imaging to prevent unintentional photochromic changes caused by the broadband visible light source.

The right panel of Figure~\ref{Vis and IR image.jpg} depicts a false-color representation of the narrowband infrared transmission contrast. The image was generated by integration of the hyperspectral imaging data of TTz-embedded polymer over the full width half maximum of the peak located at 1610~cm$^{-1}$ as indicated by the shaded area in Fig.~\ref{Transmission Spectrum}. It can be seen that the difference in dielectric function between TTz$^{2+}$ and TTz$^{0}$ states of the TTz-embedded polymer in the spectral range from 1600~cm$^{-1}$–1618~cm$^{-1}$ results in a notable contrast in the infrared transmission image. These results highlight the material’s ability to record spatial information, suggesting its applicability in tunable optical components. The ability to induce these optical changes through light exposure, without requiring high-power thermal or electrical control, positions TTz-embedded polymers as promising candidates for dynamic infrared metasurfaces.

\section{Conclusions}

This study explores the tunable optical properties of TTz-embedded polymer films for potential applications in infrared metasurfaces. The hyperspectral infrared imaging results reveal a significant photochromically-induced transmission contrast in the mid-infrared range, demonstrating the material’s capability for infrared optical modulation. Compared to conventional phase-change materials, TTz-embedded polymers could offer an alternative approach that does not require high power consumption or complex thermal control. 

The TTz-embedded polymer exhibits low absorption and a dielectric contrast in the 1500~cm$^{1}$ to 1700~cm $^{-1}$ spectral range, making it suitable for dynamic infrared light control. Unpolarized transmission measurements were conducted to further narrow down the spectral bands with sufficient dielectric contrast before and after irradiation, along with adequate transparency. The unpolarized transmission measurements reveal that the spectral range between 1500~cm$^{1}$ to 1700~cm $^{-1}$ is dominated by two prominent and narrow absorption peaks centered at approximately  1500~cm$^{1}$ and  1610~cm$^{1}$. To maximize the photochromically-induced transmission contrast during the infrared imaging experiment, the strongest absorption line located at 1610~cm$^{1}$ was selected. The difference in dielectric function between TTz$^{2+}$ and TTz$^{0}$ states of the TTz-embedded polymer in the spectral range from 1600 cm$^{-1}$–1618 cm$^{-1}$ produces a significant contrast in the infrared transmission image. These results highlight the material’s ability to record spatial information, suggesting its applicability in tunable optical components.

The ability to induce optical changes through light exposure, without requiring high-power thermal or electrical control, positions TTz-embedded polymers as promising candidates for dynamic infrared metasurfaces. The development of photochromic materials that exhibit strong and reversible changes in their optical properties in the infrared region could offer an alternative approach to achieving tunable infrared metasurfaces, potentially with advantages in terms of cost, fabrication, or power consumption. 

\authorcontributions{Conceptualization, N.~Z.~S. and T.~H.; methodology, N.~Z.~S.; software, N.~Z.~S.; validation, N.~Z.~S., T.~H. and M.~G.~W.; sample preparation, T.~J.~A, N.~F.~T;  formal analysis, N.~Z.~S.; investigation, N.~Z.~S.; resources, T.~H. G.~B., M.~G.~W.; data curation, N.~Z.~S.; writing---original draft preparation, N.~Z.~S.; writing---review and editing, N.~Z.~S.  T.~H., G.~B., N.~F.~T, T.~J.~A, D.~L., M.~G.~W.; visualization, N.~Z.~S.; supervision, T.~H., M.~G.~W.; project administration, G.~B, T.~H., M.~G.~W.; funding acquisition, T.~H. All authors have read and agreed to the published version of the manuscript.}

\funding{The authors acknowledge the support from the National Science Foundation within the IUCRC Center for Metamaterials (2052745) and National Science Foundation Grant (CHE-2400165).}

\dataavailability{ Data underlying the results presented in this paper are not publicly available at this time but may be obtained from the authors upon reasonable request.}

\acknowledgments{The authors would like to acknowledge the support from the Department of Physics and Optical Science and the Department of Chemistry at the University of North Carolina at Charlotte. We further acknowledge support from the Center for Optoelectronics and Optical Communications, the Chemistry and Nanoscale Science Ph.D. Program, the Division of Research, and the Klein College of Science at UNC Charlotte.}

\conflictsofinterest{The authors declare that they have no conflicts of interest.}


\reftitle{References}

\end{document}